\documentclass{ws-p8-50x6-00}

\begin{document}

\title{Parton saturation effects to the Drell-Yan process in the color dipole
picture}

\author{M.A. Betemps, M.B. Gay Ducati, M.V.T. Machado}

\address{Instituto de Fisica, Universidade Federal 
do Rio Grande do Sul, \\ Caixa Postal 15051, 91501-970, Porto Alegre, Brasil\\ 
E-mail: mandrebe@if.ufrgs.br, gay@if.ufrgs.br, magnus@if.ufrgs.br}




\maketitle

\abstracts{We report on the results obtained in the study
of the parton saturation effects, taken into account through the multiscattering 
Glauber-Mueller approach,  applied to the Drell-Yan (DY) process described in the 
color dipole picture. As a main result, one shows that those effects play
an important role  in the estimates of the DY differential cross section at
RHIC energies.}


The Drell-Yan (DY) process, i.e. the production of massive lepton pairs in 
hadronic collisions, in conjunction with  deep inelastic scattering 
(DIS), has been  one of the  important  processes probing  strong interaction physics. 
Recently, in connection with  the availability of high energy accelerators, a great attention has been focused on the small-$x$ region of QCD. There, the parton densities become high and the limits of the  
perturbative methods are tested. Such a region presents the onset of the saturation phenomenon (at scale $Q^2_s$), i.e. the taming of the parton (mostly gluon) distribution due to nonlinear dynamics associated with unitarity effects\cite{JBraz} .

In this contribution, we study  the high energy DY cross section 
in the target rest frame. In this case, the relevant degrees of freedom are the projectile wavefunction and the dipole-proton effective cross section. The underlying process is the scattering 
of a parton from the projectile structure function off the target color 
field. This parton radiates (bremsstrahlung) a massive photon, which subsequently decays into 
a lepton pair. The interaction with the
target can occur before or after the photon emission.  A remarkable feature
emerging is that the $\gamma^*q$-proton (or hadron) interaction can be described by  the same
$q\bar{q}$ (color dipole)-proton  cross section  as in DIS. Although diagramatically no
dipole is  present, the interference among graphs results in a product of two
quark amplitudes in the DY cross section testing the external gluonic field at
two different transverse positions (impact parameters), in a similar way to DIS\cite{Brodsky}.

In such a representation, the photoabsorbtion cross section on
deep inelastic is described by the convolution of the wavefunctions,
$\Psi_{\gamma^*}$, from the virtual photon and the interaction dipole cross
section, $\sigma_{dip}$. The wavefunctions are considered taking into
account the simplest photon  Fock state  configuration, i.e., a  $q\bar{q}$-pair fluctuation (dipole), whose transverse separation $r_{\perp}$ remains fixed during the interaction since  its  lifetime is  larger  than the collision one. The  $\sigma_{dip}$ is  modeled phenomenologically
based on a matching between  hard and soft domains, constrained  by the DIS
available data. Small dipole size configurations can be described 
through pQCD, whereas the large size ones belong to the nonperturbative
domain. Hence, one can write the photoabsorptions cross section  as a function
of the Bjorken scaling variable $\tilde{x}=x_{Bj}$, the  photon vituality scale  $\tilde{Q}^2=Q^2$ and the quark momentum fraction $\alpha$  in the quantum
mechanical form\cite{MBGMVT}, 
\begin{eqnarray} \sigma_{T,L}(\gamma^*p
\rightarrow q\bar{q}) = \int \,d^2 r_{\perp} \,\int_0^1\,d\alpha
\,\,|\Psi^{T,L}_{q\bar{q}} (\alpha
,\,r_{\perp},\,\tilde{Q}^2)|^2\,\,\sigma_{dip}(\tilde{x},\, r_{\perp})\,\,, \label{dipdis}
\end{eqnarray}
where the simplest phenomenological realization of the dipole cross section is given by the GBW model\cite{GBW}, $\sigma_{dip}=\sigma_0[1-\exp (-r_{\perp}^2 Q^2_s/4)]$, with the parameters determined by fitting small-$x$ HERA data.

In a similar way, the cross section for radiation of a virtual photon from a
quark after scattering on a proton, has the following factorized form in impact
parameter representation \cite{Kopplb},  
\begin{eqnarray}
\frac{d \, \sigma_{T,L}(qp\to \gamma^*X)}{d\ln\alpha} =\int d^2 r_{\perp}\,\,
 |\Psi^{T,L}_{\gamma^* q}(\alpha,r_{\perp},\,\tilde{Q}^2)|^2 \, \,   \sigma_{dip}(\tilde{x},\, \alpha r_{\perp}), 
\label{gplctotal}
\end{eqnarray}
where $\Psi_{\gamma^* q}$ is the quark wavefunction representing the $\gamma^* q$ fluctuation. Here, $r_{\perp}$ is the photon-quark transverse separation, $\alpha
r_{\perp}$ is the  $q\bar{q}$ separation and $\alpha$  is the fraction of 
the light-cone momentum of the initial quark taken away by the photon. Note
the difference with the DIS case, where the dipole separation is just
$r_{\perp}$ and now the dipole cross section is evaluated for  $\alpha
r_{\perp}$. The scaling variable $\tilde{x}$ is taken as $x_2=x_1-x_F$,  which are the usual invariants in the fast proton  system: $x_{1,\,2}$ are the longitudinal momentum fractions of the projectile (target) parton and $x_F$ is the Feynman variable.  We stress that such identification is far from clear in the dipole picture: for instance, one can also choose   $\tilde{x}=\alpha x_2$. In the rest frame, the quark carries momentum fraction $x_1/\alpha$ of the parent hadron, where $x_1$ is the proton's momentum fraction carried by the f\'oton. The hadronic differential cross section to Drell-Yan process is
expressed in the factorized form \cite{Kopplb},
\begin{eqnarray}
\frac{d\, \sigma^{DY}}{dM^2 \,dx_{F}}= 
\frac{\alpha_{\rm{em}}}{6\,\pi M^2}\,
\frac{1}{(x_{1} + x_{2})}\int_{x_1}^{1}\frac{d\alpha}{\alpha}
F_2^p\left(\frac{x_1}{\alpha},\,\tilde{Q}^2 \right)
\frac{d\sigma(qp \rightarrow \gamma^* X)}{d\ln\alpha}
\label{dycs}
\end{eqnarray}
where the  factor $\alpha_{\rm{em}}/(6\,\pi M^{2})$ is due to the photon decay into a lepton pair, coming from electrodynamics.  The  dilepton invariant mass is $M^2$.  Our input for 
$\sigma_{dip}$  is constrained by the
standard gluon distribution from the target corrected by saturation effects
in  high energy limit, encoded in the Glauber-Mueller (GM) approach \cite{AGL}. In  Eq. (\ref{dycs}), the structure of the projectile
is described by the proton inclusive structure function  $F_2^p(x,\tilde{Q}^2)$, in which was chosen  $\tilde{Q}^2 =M^2$. Other possible identification relying on plausive arguments is $\tilde{Q}^2=(1-x_1)M^2$ \cite{Raufhep}. Our input for the dipole cross section is given by,   
\begin{eqnarray}
\sigma_{dip}^{GM}(\tilde{x},r_{\perp}) & = & \frac{\pi^{2}
\alpha_s}{3}\,\,r^{2}\,xG_{N}^{GM} (\tilde{x},\frac{4}{r^{2}_{\perp}})\,, \label{dipGM}\\
xG^{GM}(\tilde{x},Q^2) & = & \frac{2R^{2}}{\pi^{2}}\int_{\tilde{x}}^{1}\frac{dx^{\prime}}{x^{\prime}}
\int_{1/\tilde{Q}^{2}}^{1/Q_0^{2}}\frac{d r_{\perp}^{2}}{r_{\perp}^{4}}\,\left(
\gamma _E +\ln [ \kappa_{G}  (x^{\prime}
,r_{\perp}^{2})] + E_{1}[\kappa_{G}(x^{\prime},r_{\perp}^{2})] \right) \,,\nonumber
\end{eqnarray} 
where $\gamma _E$ and $E_1(x)$ are the Euler constant and the
exponential integral. The packing factor is 
$\kappa_G=(3\pi\alpha_s r_{\perp}^2/2R^2)\,xG^{\rm{DGLAP}}$. The saturation scale $Q_s^2$ is defined through the relation  $\kappa_G(x,Q^2_s)=1$. The target transverse size is $R$, which was taken to be  $R^2=5$ GeV$^{-2}$ (strong unitarity corrections) supported by previous studies\cite{MBGMVT}. The $xG$ was parametrized by the GRV94 gluon density. 

In a recent work\cite{DYdipprd}, we have calculated the DY cross section for the available small-$x_2$ data, and found that the dipole picture in that kinematical region ( $x_2\geq 10^{-2}$) is being  tested in its limit of applicability. As an improvement,  we have introduced a reggeonic piece to account for the large $x_2$ range in the phenomenological form $\sigma_{R}(x,r_{\perp})=N_{R}\,\, r_{\perp}^2 x^{0.4525}(1-x)^3$, which supplement Eq.(\ref{dipGM}). Towards smaller $x_2$, such  contribution turns out to be negligible. In Ref.\cite{Raufhep}, the procedure was to correct the saturation scale by a threshold factor  $Q^2_s \rightarrow Q^2_s\, (1-x)^5$, and the numeric equivalence between the dipole picture (using GBW model) and NLO DGLAP calculations is shown\cite{Raufhep}, relying on an educated guess for the scales $\tilde{x}$ and $\tilde{Q}^2$.

\begin{figure}[t]
\epsfxsize=20pc 
\centerline{\epsfbox{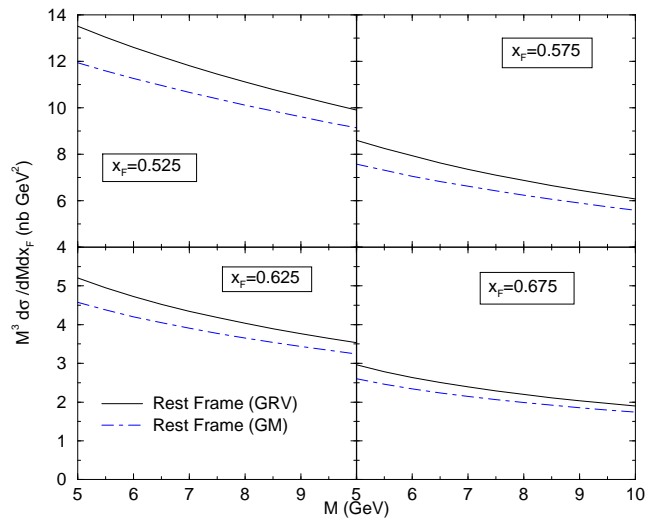}} 
\caption{The DY differential cross section per nucleon  versus $M$
for the RHIC energies ($\sqrt{s}=500$ GeV) at fixed $x_F$ in $pd$
reaction. The solid line are GM estimates, whereas  dot-dashed ones are without  gluon saturation  calculation.} 
\label{dyrhic}
\end{figure}

In order to address the color dipole picture at high energies\cite{DYdipprd} (small $x_2$), 
the DY differential cross section
for RHIC energies, $\sqrt{s}=500$ GeV,  is shown in Fig.
(\ref{dyrhic}) for the same fixed $x_F$.  In this case,  $x_2$ reaches
values of order $10^{-4}$ and unitarity effects are important. The
solid lines are the GM estimates, Eq. (\ref{dipGM}),  and the dot-dashed lines  are
the rest frame calculations with DGLAP gluon distribution (without parton saturation effects).  The
GM results  underestimates those ones without parton saturation, showing important role of the unitarity effects.  Moreover, we have obtained a 
result numerically equivalent to the LO Breit frame calculations at the RHIC energies, suggesting consistency between  both approaches. From the plots one
verifies that the deviations due to saturation effects are larger as $M$
diminishes, corresponding to smaller $x_2$ values. In absolute values, the corrections 
at RHIC energies  reach up to $\approx 20$ \%, showing that saturation effects are important in a reliable description of the observables. 
 
\section*{Acknowledgments}
\vspace{-0.3cm}
 The authors enjoyed the  lively  scientific atmosphere of this meeting. This work is supported by CNPq, Brazil. MVTM thanks Jorg Raufeisen for  discussions.

\end{document}